\documentclass[vecphys]{svmult}

\usepackage{graphicx}        
\usepackage{multicol}        
\usepackage[bottom]{footmisc}

\newcommand {\beq} {\begin{equation}}
\newcommand {\eeq} {\end{equation}}
\newcommand {\bea}{\begin{eqnarray}}
\newcommand {\eea} {\end{eqnarray}}
\newcommand{\None}{${\cal N}=1\, $}
\newcommand {\eqref} [1] {(\ref {#1})}

\begin{document}

\title*{Planar Equivalence 2006\\
 \mbox{} {\small 
A Mini-Review in Honor of Gabriele Veneziano's 65$^{\rm th}$ Birthday}}

\author{Adi Armoni\inst{1}\and
M. Shifman\inst{2}}
\institute{Department of Physics, Swansea University,
 Singleton Park, Swansea, SA2 8PP, UK
\texttt{A.Armoni@swansea.ac.uk}
\and William I. Fine Theoretical Physics Institute,
 University of Minnesota, Minneapolis, MN 55455, USA \texttt{shifman@umn.edu}}
%
%
\maketitle

\begin{abstract}
Planar equivalence between supersymmetric Yang--Mills theory
and its orientifold daughters is a promising tool for explorations of
nonperturbative aspects of quantum chromodynamics. Taking our 2004 review
as a starting point we summarize some recent developments in this issue.
\end{abstract}

\centerline{\includegraphics[width=1.7in]{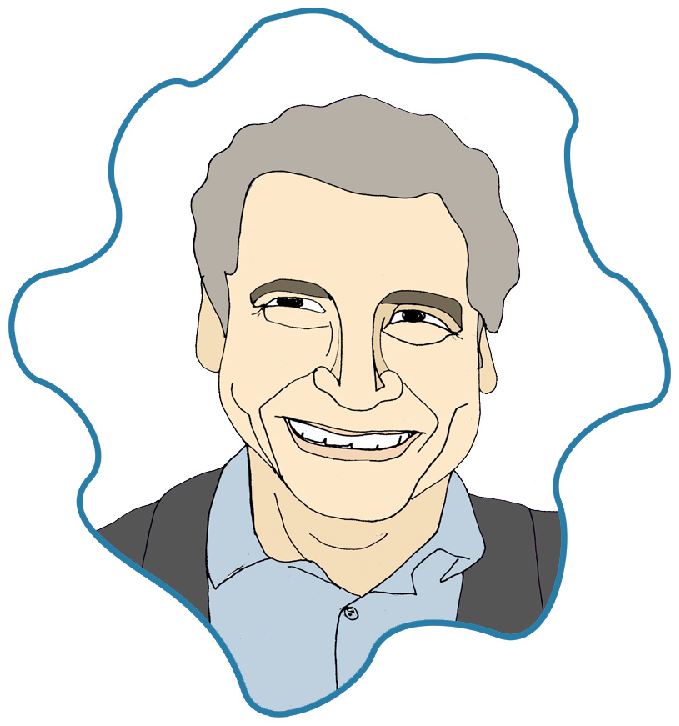}}

\mbox{}

The most interesting processes in quantum chromodynamics are 
those occurring at large distances, at strong coupling. The large distance 
dynamics determining such salient features as
chiral symmetry breaking and color confinement are the realm of nonperturbative phenomena. Despite the practical importance of the issue and the fact that this is a 
very deep theoretical problem, very few analytic methods of calculations (of a limited scope) were developed over the years, for a recent review see \cite{MS}.

The situation is much better in supersymmetric (SUSY) theories: certain quantities 
(which go under the name of $F$ terms) can be calculated {\it exactly},
due to holomorphic dependences on various parameters. In particular, it is possible to calculate the exact value of the gluino condensate \cite{VS} in pure 
\mbox{\None} super-Yang--Mills (SYM)
theory (we will also refer to this theory as supersymmetric gluodynamics).  

The basic idea behind planar equivalence is to approximate QCD by a supersymmetric theory!

The history of planar equivalence is as follows. In 1998, soon after the seminal AdS/CFT paper of Maldacena \cite{Maldacena:1997re}, Kachru and Silverstein 
\cite{Kachru:1998ys} suggested a class of nonsupersymmetric large-$N$ conformal gauge theories. The candidate theories were 
the duals of $AdS _5 \times S^5 / \Gamma$ and, therefore, named ``orbifold field theories.'' Although it turns out that these theories are in fact not conformal (not even in perturbation theory, see \cite{Armoni:2003va,Dymarsky:2005nc}) 
Kachru--Silverstein's  conjecture led to a more subtle conjecture by 
Strassler \cite{Strassler:2001fs}. A refined version of Strassler's conjecture is 
{\em planar equivalence for orientifold field theories.} In contrast to various other conjectures, the latter can be proven \cite{Armoni:2003gp,Armoni:2004ub} under rather mild assumptions, see 
Sect.~\ref{charge}. The orientifold daughter of SUSY gluodynamics is a
nonsupersymmetric Yang--Mills theory with one Dirac fermion in the two-index antisymmetric (or symmetric) representation of SU($N$). 
In this mini-review written on the occasion of Gabriele Veneziano's 65$^{\rm th}$
birthday we focus on recent developments 
in the issue of planar equivalence --- those that took place after our detailed review on this subject  \cite{Armoni:2004uu} was published.

\vspace{2mm}

The statement of planar equivalence for (the minimal) orientifold field theory is 
as follows: {\em at large $N\!,$ in a certain well defined bosonic sector, SU($N)$ 
\None SYM theory is equivalent to an SU($N)$ gauge theory with a Dirac fermion 
in the two-index antisymmetric representation.} The same statement holds for 
Dirac fermions in the two-index symmetric representation.

Although planar equivalence is an extremely interesting theoretical statement 
{\em per se}, its practical importance goes far beyond since it relates a 
supersymmetric gauge theory to a nonsupersymmetric one. Thus, potentially,  
it is a very useful tool for QCD. Let us make a simple observation: for 
SU(3)$_{\rm color}$ a Dirac fermion in the antisymmetric representation 
is equivalent to a Dirac fermion in the fundamental representation. Therefore the 
SU(3) version of the orientifold field theory  is in fact one-flavor QCD! Thus, we can approximate one-flavor QCD by supersymmetric Yang--Mills and in this way 
evaluate some nonperturbative quantities in QCD. In particular, planar equivalence 
will enable us to calculate the quark condensate in one-flavor QCD by using the 
value of the gluino condensate in supersymmetric gluodynamics.
  
\section{Planar equivalence: a refined proof}
\label{proof}

Originally, the idea of planar equivalence between supersymmetric
gluodynamics and its orientifold daughter was formulated in 2003.
Since then we refined the proof and made it more rigorous 
\cite{Armoni:2004ub}. Let us briefly outline the main ingredients of the proof.

\begin{figure}
\centering
\includegraphics[width=7cm]{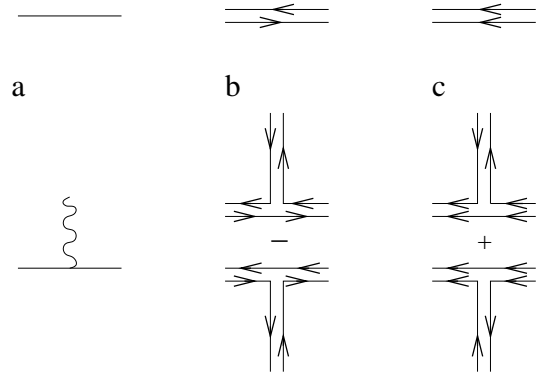}
\caption{a. The quark-gluon vertex; b. In \None SYM theory; c. In the orientifold 
field theory.}
\label{fey1}       
\end{figure}

It is instructive to start from a perturbative analysis. We want to show that all 
planar graphs of the two theories coincide. To this end it is useful to use 
't Hooft's notation. In this notation the adjoint representation is denoted by 
two parallel lines with color flow arrows pointing in the opposite directions, 
whereas the antisymmetric (symmetric) representation is denoted by 
two parallel lines with the arrows pointing in the same direction. The Feynman 
rules of the two theories are depicted in Fig.~\ref{fey1}.   

\begin{figure}
\centering
\includegraphics[width=10cm]{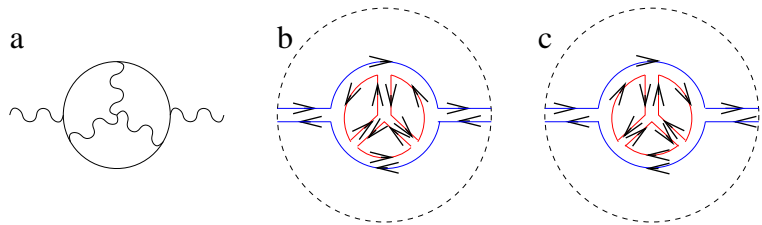}
\caption{A typical planar graph in SYM and the orientifold field theory.}
\label{fey2}       
\end{figure}

Next, we observe that the direction of the color flow arrows does not affect the 
value of the planar graphs under consideration.  To see that 
this is indeed the case, imagine that we paint every pair of the fermionic 
lines in blue and red colors, respectively. Accordingly, the gluon lines will 
be either both red or both blue. A planar graph then will be divided into 
blue regions and red regions  separated by fermionic loops. A typical example 
is given in Fig.~\ref{fey2}.

Now imagine that we reverse the arrows attached to the red lines. In this way we map a planar graph of one theory onto a planar graph of the other theory. This action does not change the value of the graph. {\em Quod Erat Demonstrandum}.  

The complete nonperturbative proof \cite{Armoni:2004ub}  is more involved, of course. The main ingredients are as follows. First, define, for a generic Dirac fermion in the representation $r$, the generating functional
\beq
\label{defW}
e^{-{\cal W}_r(J_{\rm YM},\,J_\Psi)} =  \int \, DA_\mu \, D\Psi\,  D\bar{\Psi} \, 
e^{ -S_{\rm YM} [A,J_{\rm YM} ]}\, 
\exp \left\{ \bar{\Psi} \left (i  \not\! \partial + \not\!\!  A ^a \, T^a _r + J_\Psi \right){\Psi}\right\}.
\eeq
Next,
integrate out fermions to arrive at
\beq 
\label{fint}
e^{-{\cal W}_r(J_{\rm YM},\,J_\Psi)}= \int \, DA_\mu\,
 e^{ -S_{\rm YM}[A,J_{\rm YM} ] + \Gamma_r [A, J_\Psi]}\,,
\eeq
where 
\beq
\label{defGa}
 \Gamma _r [A, J_\Psi]=\log \, \det \left( i\not\! \partial + \not\!\!  A ^a
\, T^a _r + J_\Psi\right) \, . 
\\
 \eeq
For what follows it is convenient to 
 write the effective action $\Gamma _r [A, J_\Psi]$ in the world-line
formalism \cite{wlf}, as an
integral over (super-)Wilson loops
\bea
\label{wlineint}
 \Gamma _r [A, J_\Psi] &=&
-{1\over 2} \int _0 ^\infty {dT \over T} 
\nonumber\\[3mm]
 &\times&
\int {\cal D} x {\cal D}\psi
\, \exp 
\left\{ -\int _{\epsilon} ^T d\tau \, \left ( {1\over 2} \dot x ^\mu \dot x ^\mu + {1\over
2} \psi ^\mu \dot \psi ^\mu -{1\over 2} J_\Psi ^2  \right )\right\} 
\nonumber \\[3mm]
 &\times &  {\rm Tr }\,
{\cal P}\exp \left\{   i\int _0 ^T d\tau
\,  \left (A_\mu ^a \dot x^\mu -\frac{1}{2} \psi ^\mu F_{\mu \nu} ^a \psi ^\nu
\right ) T^a _r \right\}  \, .
\eea

Thus, the generating functionals of   theories with matter in the antisymmetric/adjoint
are very similar. The dependence on the representation enters through the Wilson loops. The latter can be written as follows:
\bea
W_{\rm AS} &=&  {1\over 2} \left( ({\rm Tr}\, U)^2 - {\rm Tr}\,
U^2 \right)\, +( U\, \rightarrow U^\dagger) \label{halfAS}
, \\[3mm]
  W_{\rm adjoint} &=& {\rm Tr}\, U\, {\rm Tr}\, U^\dagger -1 +
( U\, \rightarrow U^\dagger ) = 2  \left ({\rm Tr}\, U\, {\rm Tr}\, U^\dagger -1\right) \, ,
\label{ADJ}
\eea
where  $U$ (respectively $U^\dagger$) represents the same group element in the  
 {\em fundamental} (respectively {\em antifundamental}) representation of SU($N)$. 

To complete the proof \cite{Armoni:2004ub}, one must show that at large $N$ 
one can use
\bea
W_{\rm AS} & \sim &  {1\over 2}  ({\rm Tr}\, U)^2 + {1\over 2} ({\rm Tr}\, U^\dagger)^2 \, , \\[3mm]
 W_{\rm adjoint} & \sim &  2  {\rm Tr}\, U\, {\rm Tr}\, U^\dagger \, ,
\label{ADJ2}
\eea
 and that $U$ can be replaced by $U^\dagger$ everywhere.\footnote{See 
Sect.~\ref{charge} for a more detailed discussion.} The factor $2$ in \eqref{ADJ2} is canceled by the factor ${1\over 2}$, since the adjoint representation is realized by the Majorana rather than Dirac fermions. 

A remarkable consequence of nonperturbative planar equivalence is that
(non-SUSY) orientifold field theories exhibit some feature of supersymmetric 
theories. This is surprising since the spectrum of the large-$N$ theory consists of bosons only ---
it is impossible to form finite-mass fermionic color singlets.
As a ``remnant" of SUSY they are predicted to have an even/odd parity degeneracy,
as in   supersymmetric gluodynamics. 
More generally, two bosons from one and the same 
would-be supermultiplet, must be degenerate in mass at $N\to\infty$.
In addition, the quark condensate $\langle \bar \Psi \Psi \rangle $ will form, and its value will be identical to that of the gluino condensate in \None SYM theory. Other important properties are the NSVZ $\beta$ function, the domain wall spectrum and 
gluonic Green functions \cite{Armoni:2003gp,Armoni:2004uu}.

\section{The orientifold large-$N$ expansion}
\label{expansion}

Let us forget for a short while about supersymmetry and look at 
planar equivalence from a broader perspective. Assume we are 
interested in the large-$N$
limit of multiflavor QCD. 
There are various options of generalizing SU(3)  QCD to SU($N)$ gauge theory. 
In the original 't Hooft large-$N$ expansion \cite{GH} both $g^2 N$ and the number of flavors $N_f$ (quarks in the fundamental representation)
is kept fixed (this is realized in the modern gauge/string duality by keeping the number of flavor branes fixed). In the Veneziano large-$N$ expansion (the topological expansion \cite{TE}) the ratio $N_f/N$ is kept fixed, together with $g^2N$ (it can be achieved by placing branes on orbifold singularities, in a certain region of the moduli space). The advantage of the latter expansion is that the quark loops are not suppressed at
large $N$ and, hence,  flavor physics is better captured in this approximation. 
In particular, the $\eta '$ mass does not vanish even when $N \rightarrow \infty$, that is to say,  a massive $\eta '$ is a part of the planar theory.

While both expansions are interesting and useful,  
there is no full quantitative solution to either. It is tempting to say that large-$N$ QCD is dual to a string theory, and there was a significant progress along these lines \cite{Maldacena:1997re}, but it would be certainly wrong to say
that an accurate and well-developed description of QCD has been already attained.
Therefore, alternative large-$N$ limits may well prove to be very useful.

\begin{figure}
\centering
\includegraphics[height=1cm]{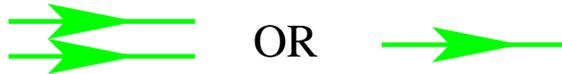}
\caption{Antisymmetric/fundamental representation in SU(3).}
\label{or}       
\end{figure}

Let us discuss a new {\em orientifold} large-$N$ expansion \cite{Armoni:2003fb}. It will lead to certain {\em quantitative} predictions for QCD. We start from  
SU(3)$_{\rm color}$ Yang--Mills theory with $N_f$ 
quark flavors in the fundamental representation (to be referred to 
as multiflavor QCD). Since for SU(3) the Dirac fermion in the fundamental 
representation is equivalent to the Dirac fermion in the antisymmetric 
two-index representation, we have the option of generalizing the theory to 
SU($N)_{\rm color}$  treating $N_f$ fermions as antisymmetric Dirac fermions, see 
Fig.~\ref{or}. 

The next step is to consider the large-$N$ limit of this theory while keeping 
$N_f$ fixed. This large-$N$ approximation, to be referred to 
as the {\em orientifold large}-$N$ approximation, is somewhat similar to the topological expansion since the quark loops are not suppressed with respect to the gluon loops. 

Through planar equivalence the theory with $N_f$
Dirac quarks in the two-index antisymmetric approximation
is related to the theory with $N_f$ adjoint Majorana quarks
(in the common sector).

While phenomenological consequences of the orientifold
large-$N$ limit so far remain essentially unexplored, 
in purely theoretical aspect planar equivalence 
of these theories revived interest in gauge theories with
quarks in higher representations, other than the fundamental representation.
In particular one can ask the question as to the form of the chiral Lagrangian
in the Yang--Mills theory with antisymmetric or adjoint quarks.
The chiral Lagrangian of QCD with fundamental quarks
supports Skyrmions which can be identified with baryons
\cite{EW}. And what about the chiral Lagrangian in the theory
with antisymmetric quarks?
The pattern of the spontaneous breaking of the chiral symmetry
in this {\em gedanken} case is well-known. The corresponding chiral Lagrangian is not drastically different from that of QCD. It supports Skyrmions too.
However, the mass of the  Skyrmions in this case scales as $N^2$
rather than $N,$ as is the case in the 't Hooft limit. 
At first sight, there is no apparent match between Skyrmions and baryons. 
It turns out \cite{B} that $N$-quark hadrons built of antisymmetric fermions
are unstable with regards to fusion of $N$ species into a huge
compound object built of $N^2$ quarks. It is the latter which is an analog of 
the baryon! For subsequent discussions see \cite{CC}.

Moreover, chiral Lagrangians were found in theories with the adjoint quarks
\cite{abs}. The issue of baryon analogs and Skyrmions in this case is intriguing and subtle. There is no conservation of fermion number; rather it is
$(-1)^F$ which is conserved. It was argued \cite{abs}
that an analog of the baryon is a compound object built of $N^2$ quarks
with an abnormal assignment of $(-1)^F$. On the Skyrmion side
it is seen as a Hopf Skyrmion
whose topological stability is associated with a nontrivial Hopf invariant.

\section{Applications for one-flavor QCD}
\label{oneflavor}

As we explained in   Sects.~\ref{proof}  and \ref{expansion}, we can approximate 
one-flavor QCD by a planar theory with one Dirac two-index antisymmetric fermion. This theory is planar-equivalent to \None SYM theory. We can therefore make several 
{\em quantitative} predictions about the nonperturbative regime of the one-flavor QCD.

The first prediction concerns the spectrum of the theory. As we discussed at the end of Sect.~\ref{proof}, the color-singlet spectrum of the orientifold field theory
exhibits an odd/even parity degeneracy. Thus, we expect a similar degeneracy in the spectrum of one-flavor QCD, within a $1/N$ error,

\beq
{M ^S_- \over M ^S _+} = 1 +{\cal O}(1/N)\, ,
\eeq 
where $M^S_-$ is a color-singlet bosonic degree of freedom with spin $S$ and {\em odd} parity and $M^S_+$  is a color-singlet bosonic degree of freedom with spin $S$ and {\em even} parity. In particular the $\eta'$ and the $\sigma$ mesons should be approximately degenerate. This prediction was supported by  lattice QCD
analyses, see Ref.~\cite{Keith-Hynes:2006wm}.

Another prediction is the value of the quark condensate in one-flavor QCD. The analysis  carried out in \cite{Armoni:2003yv} was recently tested in a lattice simulation by DeGrand et al.~\cite{DeGrand:2006uy}.
A comment on this issue is in order here.
It is convenient to deal with a renormalization-group
invariant definition of the gluino condensate and the quark condensate,
\beq 
\langle  \bar \Psi \Psi \rangle _{\rm RGI} \equiv (g^2)^{\gamma / \beta} \langle  \bar \Psi \Psi \rangle  \,.
\eeq
The renormalization-group
invariant value of the gluino condensate is
\beq
\langle \lambda \lambda \rangle = -{N^2 \over 2\pi ^2} \Lambda ^3 \, .
\eeq
Nonperturbative planar equivalence implies the equality of the orientifold quark condensate and the gluino condensate at infinite $N$. Moreover, since we know that for $N=2$ the antisymmetric representation is equivalent to a color-singlet we can make an educated guess that the value of the quark condensate at any $N$ is
 \beq
\langle \bar \Psi \Psi \rangle = -\left (1-{2\over N} \right ){N^2 \over 2\pi ^2} \Lambda ^3 \, .
\eeq 
The evaluation of the quark condensate for $N=3$ (one-flavor QCD) at $2$ GeV (assuming the 't Hooft coupling is 0.115) yields
 \beq
\langle \bar q q  \rangle ^{\rm orientifold} _{\rm 2\,\, GeV} = - (262\,\,{\rm MeV})^3 \,\, \pm 30\%\,. \label{onefresult}
\eeq
This value can be compared with a recent lattice evaluation by DeGrand 
et al.~\cite{DeGrand:2006uy}
\beq 
\langle \bar q q  \rangle ^{\rm lattice} _{\rm 2\,\, GeV} = - (269(9)\,\,{\rm MeV})^3 \,.
\eeq
The agreement is more than satisfactory.

\section{Applications for three-flavor QCD}
\label{threeflavors}
\noindent 

Is it possible to use planar equivalence to calculate nonperturbative quantities in real  three-flavor QCD?
In a bid to answer this question positively,
a ``mixed'' approach has been suggested.

 Consider an SU($N)$ gauge theory with one Dirac fermion $\Psi$ 
  in the antisymmetric representation and two extra Dirac fermions $\chi ^i$ in the fundamental representation. For SU(3) this model reduces to three-flavor QCD. When $N \rightarrow \infty$ the fundamental flavors can be neglected and our model is planar equivalent to \None SYM theory. Thus, the model at hand interpolates between QCD for  SU(3)  and SYM theory at large $N$.

Several subtleties arise while considering this model. Because of 
a chiral symmetry breaking   Goldstone bosons occur in this model, at any finite $N$. Therefore, in the attempt to match quantities of this theory and \None SYM theory, one has to choose sources which do not couple to these Goldstone particles. 

A detailed analysis of the model \cite{Armoni:2005wt} leads to the estimate
\beq
\langle \bar \Psi \Psi \rangle _{\rm RGI} / \Lambda ^3 = -\left ( 1- {2\over N} \right ){N^2\over 2\pi ^2} \, ,
\eeq
just as in the previous case. Note, however, that in this model $\beta_0 = 3N$ (as in SYM theory), and, as a result, the running coupling is different than in one-flavor QCD. As a result, we find, instead of \eqref{onefresult}
\beq
\langle \bar qq \rangle ^{\rm orientifold}_ {\rm 2\, GeV} ~~=~~ -(317 ~\pm~ 30~\pm~36~ {\rm MeV})^3 \, .
\eeq
The errors here are due to the 30\% uncertainty of the $1/N$ formula and the experimental uncertainty in the 't Hooft coupling at 2 GeV.
The above  prediction should be compared with a recent lattice analysis by McNeile \cite{McNeile:2005pd}
 \beq
\langle \bar qq \rangle ^{\rm lattice}_ {\rm 2\, GeV} ~~=~~ -(259 ~\pm~ 27~ {\rm MeV})^3 \, .
\eeq

\begin{figure}
\centering
\includegraphics[height=7cm]{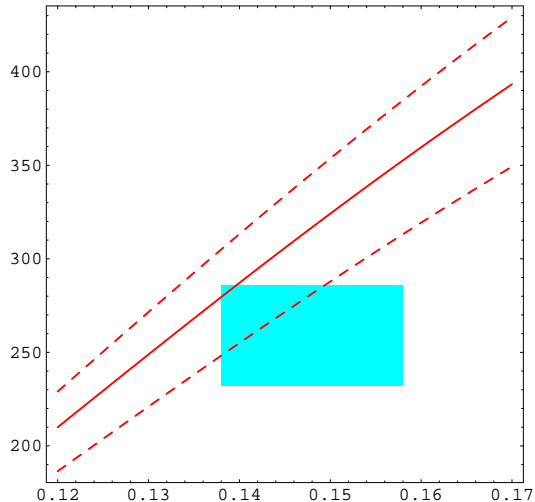}
\caption{The quark condensate expressed as $-(y~{\rm MeV})^3$ as a function of
the 't Hooft coupling $\lambda$. The solid line represent he prediction of planar equivalence. The two dashed lines represent the $\pm 30\%$ error. The $\pm 1\sigma$ range of the coupling,
$0.138 < \lambda < 0.158$ and the lattice estimate
$-(259 \pm 27~{\rm MeV})^3$ define the shaded region.}
\label{condensate}       
\end{figure}

The orientifold prediction and the lattice simulation result are confronted
in Fig.~\ref{condensate}.

\section{Sagnotti's Model and the Gauge/String Correspondence}
\label{sagnotti}

\noindent 

Orientifold field theories originate in string theory. The starting point is 10D 
type-0B string theory. By adding the orientifold 
$$\Omega' \equiv \Omega (-1)^{f_R}$$ 
and 32 D9 branes we end up with a nonsupersymmetric nontachyonic 
string theory \cite{Sagnotti:1995ga,Sagnotti:1996qj}. The low-energy spectrum of the closed string modes consists of the dilaton, the graviton, and a set of the 
Ramond--Ramond (RR) fields. There are no fermions (the Neveu--Schwarz--Ramond sector). The open string sector consists of a ten-dimensional $U(32)$ gauge theory with an antisymmetric fermion. The model is free of RR tadpoles.

In order to obtain a realization of the 4D orientifold field theory one can use a 
Hanany-Witten brane configuration in type 0A, namely a set of $N$ D4 branes and O'4 plane suspended between rotated NS5 branes \cite{Armoni:2003gp}. An alternative realization \cite{DiVecchia:2004ev} is via fractional D3 branes placed on a 
$C^3 / Z_2 \times Z_2$ orbifold singularity in type 0'B. The latter description 
is useful for the gauge/gravity correspondence \cite{Armoni:2005qr}. 
Since at $g_{\rm st}=0$ the bosonic gravity modes of type 0'B and their interactions are identical to those of type IIB, the gauge/gravity correspondence (provided that it holds) provides an additional evidence in favor of planar equivalence: if the bosonic sectors of two gauge theories are described by the same bosonic sectors of two string theories at $g_{st}=0$ then the two gauge theories must be equivalent at infinite $N$.

The gauge/gravity correspondence for the orientifold field theories was used
recently \cite{Armoni:2005qr} to make predictions regarding the theories at finite
$N$. In contrast to the supersymmetric type IIB background which contains $N$ units of the RR flux, the type-0B background contains $N-2$ units of the RR flux, due to the presence of the $O'5$ plane that shifts the flux by $-2$. Certain quantities are sensitive to this shift. This is in agreement with results from 
the effective action approach presented in \cite{Sannino:2003xe}.

\section{Charge Conjugation and the Validity of Planar Equivalence}
\label{charge}

\noindent 

Recently it was pointed out \cite{Unsal:2006pj} that a necessary and {\em sufficient} condition for orientifold planar equivalence to hold is 
the absence of the spontaneous breaking of charge conjugation symmetry
(for earlier  work related to planar equivalence between SYM theories and 
{\em orbifold}
daughters see Ref.~\cite{Kovtun:2004bz}).
This assumption was implicit in our refined proof
\cite{Armoni:2004ub}. It is clear that this issue deserves
a separate discussion in Yang--Mills theory {\em per se},
not necessarily in association with supersymmetry or planar equivalence.

Motivated by \cite{Unsal:2006pj} we argued \cite{Armoni:2007rf} that $C$ parity does not break spontaneously in any vector-like gauge theory on $R^4$. We first argued that charge conjugation is not broken in pure Yang--Mills theory. Our reasoning is based on the uniqueness of the Yang--Mills vacuum. 
Being physically compelling our arguments, unfortunately, stop short of
a rigorous mathematical proof of the type given in \cite{VW} 
regarding $P$ parity. There is a deep distinction between these two aspects of QCD.
While the spatial parity conservation is essentially nondynamical
and is based on a general feature of vector-like gauge theories
with spinor quarks, the $C$-parity conservation versus nonconservation
is a dynamical question. The uniqueness of the Yang--Mills vacuum
provides us with the necessary dynamical information.

Then we prove  \cite{Armoni:2007rf} that if the charge conjugation is unbroken in pure Yang--Mills it is not broken in any vector-like theory.

The above arguments are general and apply to QCD as well as to any other vector-like theory. The absence of the spontaneous breaking of $C$ parity is sufficient for planar equivalence to be valid.  It is instructive to return to  the proof \cite{Armoni:2004ub} and heck where exactly we assume charge conjugation to hold. 

In fact, as was noted in Sect.~\ref{proof}, we need to assume the expectation values of traces of all Wilson loops to coincide with those of their conjugated, being evaluated in the pure Yang--Mills vacuum. 
This requires unbroken $C$ parity of pure Yang--Mills theory.
Once it is established, it automatically covers the theories with vector-like quarks provided that the expansion in quark loops is convergent.

\section{Other developments}
\label{recent}

\noindent 

Planar equivalence was used in both formal works and in phenomenology. Papers on the subject appeared on all theoretical high-energy archives: hep-th, hep-ph and hep-lat. Here
we would like to mention a few.

The lattice works are mainly devoted to verification of planar equivalence. A formal strong-coupling and large mass proof was given by Patella~\cite{Patella:2005vx}. The paper by DeGrand et al.~\cite{DeGrand:2006uy} confirms our prediction for the quark condensate in one-flavor QCD. The prediction regarding the mass ratio 
$m^2 _{\eta '}/ m^2 _{\sigma}$ was confirmed by Keith-Hynes and Thacker 
\cite{Keith-Hynes:2006wm}.

Phenomenological papers, mainly by Sannino and collaborators 
\cite{Sannino:2004qp,Hong:2004td} were devoted to
constructing of technicolor models based on the orientifold field theories with symmetric matter. In another recent work \cite{Strassler:2006im}, predictions about one-flavor QCD were used for ``beyond the standard model phenomenology.''

Among the more formal aspects, it is worth mentioning the work by
 Di~Vecchia et al. \cite{DiVecchia:2004ev} who studied realizations of the orientifold field theories in type-0' string theory as well as tree level string amplitudes in these models. 

A partial list of other related works is given in Refs.~\cite{Feo:2004mr,Barbon:2005zj,Veneziano:2005qs,DeGrand:2006qb,Hollowood:2006cq}.

Summarizing, planar equivalence is a new useful tool 
in a very limited toolkit available at present for
calculations of nonperturbative quantities in QCD. It has already resulted 
in a few promising applications, both in QCD, string theory, AdS/CFT, lattice gauge theory and beyond the standard model phenomenology. We believe that further studies are needed in order to exploit the potential of this method. In particular, it seems
promising to search for new 
planar-equivalent pairs with the aim of learning about one of them from the other.

\section*{Acknowledgments}

We are happy to thank Gabriele Veneziano for a fruitful and enjoyable collaboration.
We are grateful to Courtney Davis for the kind permission
to use her cartoon of Gabriele Veneziano from {\sl (M)agazine}$^{\,\footnotesize\rm one}$ 2006.

A.A.~is supported by the PPARC advanced fellowship award.
The work of M.S. is
supported in part by DOE grant DE-FG02-94ER408.

\end{document}